\def\IR{\relax{\rm I\kern-.18em R}}
\def\IN{\relax{\rm I\kern-.18em N}}

\def\C{{\cal C}}
\def\l{\lambda}

\def\IZ{\relax\ifmmode\mathchoice
{\hbox{\cmss Z\kern-.4em Z}}{\hbox{\cmss Z\kern-.4em Z}}
{\lower.9pt\hbox{\cmsss Z\kern-.4em Z}}
{\lower1.2pt\hbox{\cmsss Z\kern-.4em Z}}\else{\cmss Z\kern-.4em Z}\fi}

\newcommand{\be}{\begin{equation}} \newcommand{\ee}{\end{equation}}
\newcommand{\bea}{\begin{eqnarray}} \newcommand{\eea}{\end{eqnarray}}

\documentstyle[aps,prl,multicol]{revtex}
\begin{document}

\title{Relative entropy in $2d$ Quantum Field Theory, 
finite-size corrections and irreversibility of the Renormalization Group
}

\author{Jos{\'e} Gaite}
\address{Instituto de Matem\'aticas y F{\'\i}sica Fundamental, 
             CSIC, Serrano 123, 28006 Madrid, Spain}

\maketitle

\begin{abstract}
The relative entropy 
in two-dimensional Field Theory is studied
for its application as an irreversible quantity under the Renormalization 
Group, relying on a general 
monotonicity theorem for that quantity previously established. 
In the cylinder geometry, interpreted as finite-temperature field theory, 
one can define from the relative entropy a monotonic quantity similar to 
Zamolodchikov's $c$ function. 
On the other hand, the one-dimensional quantum thermodynamic entropy also 
leads to a monotonic quantity, with different properties. 
The relation of thermodynamic quantities with the complex 
components of the stress tensor is also established and hence 
the entropic $c$ theorems are proposed 
as analogues of Zamolodchikov's $c$ theorem for the cylinder geometry.
\end{abstract}
\pacs{11.10.Gh, 05.70.Jk, 11.10.Kk}
\begin{multicols}{2}
In Euclidean 
Quantum Field Theory (QFT) it is possible to define a type of entropy, 
the relative entropy, which is a monotonic 
function of the couplings and increases in the crossover from one 
multi-critical point to another of lower order \cite{I-OC}. 
Therefore, it is a suitable 
quantity to embody the irreversibility of the action of 
the renormalization group. There is also 
a well-known and celebrated monotonic quantity in 
two-dimensional ($2d$) QFT, Zamolodchikov's 
$c$ function \cite{Zamo}. Although a priory there is no connection 
between both quantities, some arguments indicate that such a connection may 
nevertheless exist. For example, Zamolodchikov's $c$ function is supposed 
to count the independent degrees of freedom in a model near 
the critical point (CP). This certainly agrees with the statistical definition 
of entropy. 
On the other hand, the central charge of $2d$ conformal field theory (CFT),  
to which Zamolodchikov's $c$ function reduces at the CP, has 
been shown to coincide with a particular type of renormalized entropy, 
the geometric entropy \cite{HLW}. 
These arguments beg for an investigation on whether a relation between
Zamolodchikov's $c$ function and some type of entropy in $2d$ exists 
off the CP. 

We study here the properties of the relative entropy  
in general $2d$ models, in regard to its connection with Zamolodchikov's 
$c$ function, with explicit computations for the Gaussian and
the Ising models. We shall always consider continuum theories 
with UV cutoff $\Lambda$ and we shall further 
introduce an IR cutoff, for example, by giving the system a 
finite size. 
A particular finite geometry has an interesting interpretation: 
The classical partition function in the strip or cylinder of width $\beta$ 
is equivalent to the one-dimensional quantum partition function at
temperature $T=1/\beta$. We can then calculate thermodynamic functions of
this quantum system, for example, the quantum specific entropy. 
This entropy will also have a r\^ole as a monotonic quantity. 

The idea of applying finite-size scaling methods to Zamolodchikov's theorem 
appeared in \cite{BoyHo,MM}.
The form of the finite-size corrections to the free energy for 
the Gaussian model and the Ising model off their CP has been
obtained in \cite{NaOC}. There it is briefly discussed their connection
with Zamolodchikov's $c$ function, concluding that they differ but making no 
further analysis. 
The compact dimension $\beta$ can be used as 
RG parameter, providing a thermodynamic interpretation of the RG 
\cite{CaNFr,Zab}. The finite-size correction to the free energy was used 
as a candidate monotonic function in \cite{CaNFr}, 
concluding that a thermodynamic analogue of 
Zamolodchikov's theorem holds for it but only under an 
additional condition which cannot be deduced from thermodynamics rules.
Here, relying on the monotonicity theorem for the relative entropy \cite{I-OC} and an analysis of the relation of thermodynamic quantities with the complex 
components of the stress tensor, we shall propose a 
monotonicity theorem analogous to Zamolodchikov's. We shall further prove 
a new thermodynamic monotonicity theorem involving 
the quantum specific entropy.

Given the cutoff 
logarithm of the partition function per unit volume $W[\l,\Lambda]$, 
the relative entropy is the Legendre transform of
$W[\l,\Lambda]-W[0,\Lambda]$ with respect to the relevant couplings $\l_a$,
\begin{equation}
S_{\rm rel} = W - W_0 - \l^a\partial_a W. \label{Legtrans}
\end{equation}
Now we select one coupling $\l$---or take a common factor of all the couplings%
---to evaluate the change of $S_{\rm rel}$ with respect to it. 
To be precise, we must use the {\em difference} between the couplings 
and their critical values, since the CP is taken as the reference for 
the relative entropy. (The critical couplings may be null in some cases.) 
We have the following general monotonicity theorem \cite{I-OC}
\begin{equation}
\l{dS_{\rm rel}\over d\l} = 
\langle (I_\l-\langle I_\l\rangle)^2\rangle \geq 0,
\label{mono}
\end{equation}
where $I_\l = \l\int\Phi$ is the relevant part of the action 
containing the coupling 
that we consider. 
Let us introduce the stress tensor trace, $\Theta := T_a^a$, which 
in general is proportional to the relevant part of the action; 
more precisely,
$$\Theta = \l\,y\,\Phi,\quad \hbox{with}\quad y = 2 - d_{\Phi} > 0,$$ 
where $d_{\Phi}$ is the scaling dimension of the field $\Phi$.  
Hence, we can write the monotonicity theorem as
\be
\l{\partial S_{\rm rel}\over \partial \l} = {1\over y^2}\int \!\!d^2z\,
\langle[\Theta(z) - \langle\Theta(0) \rangle] 
[\Theta(0) - \langle\Theta(0) \rangle]\rangle \geq 0. 
\label{monoSrel}
\ee

We would like to derive a general expression for $S_{\rm rel}$. 
We can use the scaling form
\be
W(\l,\Lambda)= \Lambda^{2}\,{\cal F}({\l^{2\over y}\over\Lambda^{2}}).
\ee
In absence of logarithmic corrections, 
${\cal F}$ is an analytic function 
\cite{Zamores}, so $W$ can be expanded as
\be
W(\l,\Lambda)= \Lambda^{2}\,F_0 + F_1\,\l^{2\over y}+
 {\rm O}(\Lambda^{-2}).
\ee
The UV divergent term is irrelevant for the relative entropy and 
in the infinite cutoff limit 
\begin{equation}
S_{\rm rel}(\l) = W(\l) - W(0) - \l{dW\over d\l} = 
F_1\,\frac{y-2}{y}\,\l^{2\over y}.  
\label{Srel}
\end{equation}
Taking into account that $y < 2$ and $F_1 < 0$ we have that $S_{\rm rel}(\l) 
> 0$ and it increases with $|\l|$. However, given the simple scaling form of 
$S_{\rm rel}(\l)$, this statement is not very informative. We will 
obtain a more illuminating version when we introduce a finite geometry. 

Let us now consider solvable models, namely, the Gaussian model 
($d_{\Phi}=0$) and the Ising model ($d_{\Phi}=1$), which, 
on the other hand, exhibit logarithmic corrections. 
The relative entropy per unit volume of the Gaussian model calculated 
using dimensional regularization was given in \cite{I-OC}. 
It can be expressed as 
$$ S=\frac{\Gamma[(4-d)/2]}{(4\pi)^{d/2}\,d}\,t^{d/2},$$ which in $d=2$ yields
\begin{equation}
S= \frac{t}{8\pi},
\end{equation}
with $t = r - r_c = r \equiv m^2$, $m$ being the mass parameter.

It is more illustrative to start with the expression of the cutoff 
logarithm of the partition function per unit volume
\begin{equation}
W(r) \equiv -\ln Z =\pm {1\over2}\int_{0}^{\Lambda}{d^2p\over
{(2\pi)}^2}\,\ln{p^2+r\over\Lambda^2}
\label{W}
\end{equation}
for free bosons (upper sign) or Majorana fermions (lower sign). 
It can be integrated exactly and yields
\begin{equation}
W(r) = \pm{1\over 8\pi} \left\{-{\Lambda^2} + {r} \ln {\Lambda^2\over r} 
+ {r} + {\rm O}(\Lambda^{-2})\right\},
\label{Wexpa}
\end{equation}
exhibiting a quadratic and a logarithmic divergence. We have 
in the infinite cutoff limit 
\begin{equation}
S_{\rm rel}(r) = W(r) - W(0) - r{dW\over dr} = \frac{r}{8\pi},  \label{SrelG}
\end{equation}
for the Gaussian model, 
in accord with the dimensional regularization result. 
For the Ising model,
\begin{equation}
S_{\rm rel}(r) = W(r) - W(0) - m{dW\over dm} = -\frac{m^2}{8\pi}\,
(1+\ln {m^2\over \Lambda^2}).  \label{SrelI}
\end{equation}
The relative entropy is monotonic with $m^2$ for both models. However, 
the presence of the logarithmic correction in the latter case signals that 
it is not well defined in the continuum limit $\Lambda \rightarrow \infty$ 
unless we introduce a renormalization scale or, alternatively, 
an IR cutoff. This is the general situation for models with 
logarithmic corrections.

Let us now consider a finite size geometry, 
in particular, a cylinder, equivalent to finite 
temperature field theory. It provides an IR cutoff with 
physical interest. 
The partition function is $Z = {\rm Tr}\,e^{-\beta\,H}$, 
which can be represented
as a functional integral on $S^1 \times \IR$ with $\beta=1/T$ the length of 
the compact dimension.
The specific logarithm of the partition function on a cylinder 
of width $\beta$ and length $L$ has a finite-size expression as $L \to \infty$ 
\begin{equation}
{-\ln Z\over L} = \beta\,{F\over L} = 
e_0(\Lambda,m)\, \beta + {C(\beta,m)\over\beta},   \label{fse}
\end{equation}
with $C(\beta,m)$ a universal dimensionless function.
Defining $x= m\,\beta$, we can write it as a single variable
function $C(x)$. At criticality it is 
proportional to the CFT central charge, 
$C(0) =-\pi\,c/6$ \cite{BCN,Aff}. One can readily calculate the $1d$ energy
\begin{equation}
{E\over L} = -{\partial\ln Z/L\over\partial\beta} = e_0  - 
{1\over\beta^2}\left(C -
\beta{\partial C\over\partial\beta}\right) .
\label{E}
\end{equation}
From the energy we can compute the thermodynamic entropy
\begin{equation}
{S\over L} = \beta\,{E-F\over L}  = -2\,{C\over\beta} + 
{\partial C\over\partial\beta} = {\pi\,c\over 3\,\beta}+ {\rm O}(1).
\label{SwC}
\end{equation}
The specific ground state energy $e_0$ does not contribute to the entropy, 
which vanishes in the ground state, in accord with the third law
of thermodynamics. At the CP $S/L =\pi\,c/(3\,\beta)$, which 
is reminiscent of the relation between geometric entropy for a CFT and 
central charge found in \cite{HLW}.

The theorem of increase of the relative entropy (\ref{monoSrel}) holds 
in general on a finite geometry and guarantees that $S_{\rm rel}(\l,\beta)$ 
increases with $\l$ or, alternatively, with $m \propto \l^{1\over y}$. 
At the CP the theory is conformal invariant 
and $\Theta(z)=0$; hence $\l\,\partial S_{\rm rel}/ \partial \l=
(m/y)\,\partial S_{\rm rel}/ \partial m=0$. Therefore, 
we propose to define an off-critical ``central charge" 
\begin{equation}
{\cal C}(x) = \beta^2\,S_{\rm rel}(m,\beta) 
\end{equation}
which is monotonic with $x$ and plays a similar r{\^o}le to Zamolodchikov's 
$c$ function. 
Thus we can express the monotonicity theorem in terms of dimensionless 
quantities simply as
\begin{equation}
x{d {\cal C}\over d x} = 
{\beta^2\over y}\int\!\! d^2z\,\langle(\Theta(z) - \langle\Theta(0) \rangle) 
(\Theta(0) - \langle\Theta(0) \rangle)\rangle.
\end{equation}

While this form resembles that of Zamolodchikov, it is not quite 
the same. The correlator of $\Theta$'s in the second term appears integrated.
Furthermore, a detailed calculation of Zamolodchikov's function $c(m)$ 
for the free boson or fermion 
shows that it differs from ${\cal C}(m) = {\cal C}(x)|_{\beta=1}$. The
cause is actually geometrical: A crucial step in the proof of Zamolodchikov's 
theorem relies on the assumption of rotation symmetry \cite{Zamo}, 
which does not exist on 
the cylinder. Hence, one cannot obtain the theorem for it, 
contrary to the assertion in Ref.\ \cite{Zab}. 
However, the absence of rotation symmetry is traded for the appearance of a 
new parameter, the length $\beta$, which can be used to obtain the
monotonicity theorem above. 

Besides, we may consider the behaviour of the absolute $1d$ quantum 
entropy $S$ with respect to $\beta$:
\begin{equation}
{\partial S\over \partial\beta} = 
{\partial \over \partial\beta}(\beta\,E- \beta\,F) = 
\beta\,{\partial E\over \partial\beta} = 
\beta\,{\partial^2(\beta\,F)\over \partial\beta^2}.    \label{monoS}
\end{equation}
We have again monotonicity, for $\beta\,F$ is a convex function of $\beta$, as 
deduced from the expression of its second derivative as the average 
$-\langle (H - \langle H \rangle)^2 \rangle$ with $H$ the {\em total}
Hamiltonian, 
that is, including the kinetic term, unlike the monotonicity in \cite{I-OC}.
This monotonicity is in principle unrelated with the monotonicity of 
$S_{\rm rel}$ with respect to $m$.
It allows us to define another monotonic dimensionless function, 
${\tilde\C}(x) = S/(L\,m).$ 
At the CP $S/L=\pi\,c/(3\,\beta)$, implying that ${\tilde\C}(x)$ diverges at 
$x=0$, whereas ${\cal C}(0) = 0$. On the other hand, in the IR zone, 
$x \gg 1$, ${\cal C}$ diverges as ${\cal C}(x)\sim x^2$, whereas 
$\tilde\C(x)$ decays exponentially. 

We illustrate the form of finite size corrections again with solvable models. 
For the Gaussian model the correction to the free energy 
can be expressed as the free energy of an ideal Bose gas, 
\begin{equation}
\beta\,{F\over L} = e_0\, \beta + \int_{-\infty}^\infty {dp\over 2\pi}\, 
\ln\left(1- e^{-\beta\,\epsilon(p)}\right),
\label{intF}
\end{equation}
where the one-particle energy is $\epsilon(p) = \sqrt{p^2+m^2}$.
This formula can also be obtained by an explicit calculation of 
the finite-size corrections \cite{NaOC}. When $m=0$ it can be used to
calculate the central charge \cite{Aff}. However, an expansion in
powers of $m$ is not advisable: The ensuing integral at the next order is
IR divergent; that is to say, the expression (\ref{intF}) is non analytic
at $m=0$. Fortunately, the integral
can be computed by changing the integration variable to $\epsilon$ and
expanding the logarithm in powers of $e^{-\beta\,\epsilon}$. 
One obtains
\begin{equation}
\beta\,{F\over L} = e_0\, \beta  
-{m\over\pi} \sum_{n=1}^\infty {1\over n}\,K_1(n\,m\,\beta), \label{FK}
\end{equation}
where $K_1(x)$ is a modified Bessel function of the second kind. 
For large $x= m\,\beta/(2\pi)$
the correction is exponentially negligible but 
a small-$x$ expansion yields
\begin{eqnarray}
{C(x)\over {2\pi}} = -{\zeta(2)\over2\pi^2} + {x\over 2} +
{x^2\over 2}\left(\ln{x\over 2}+\gamma-{1\over 2}\right) + \nonumber\\
\sum_{l=2}^{\infty} \left(\!\!\begin{array}{c}1/2\\l\end{array}\!\!\right) 
x^{2l}\, \zeta(2l-1).
\label{FG}
\end{eqnarray}
The first term
$\zeta(2) = \pi^2/6$ gives the usual critical part and central charge $c=1$. 
The specific entropy (\ref{SwC}) is 
\begin{equation}
{S\over L} = {\pi\over 3\,\beta} - {1\over 2}\,m + \beta\,{m^2\over 4\pi} 
+ {\rm O}(m^4).
\label{SG}
\end{equation}
 
For the Ising model we have instead an ideal Fermi gas, 
\begin{eqnarray}
\beta\,{F\over L} &=& e_0\, \beta -\int_{-\infty}^\infty {dp\over 2\pi}\, 
\ln\left(1+ e^{-\beta\,\epsilon(p)}\right) \nonumber\\ &=& e_0\, \beta + 
{m\over\pi} \sum_{n=1}^\infty {(-)^n\over n}\,K_1(n\,m\,\beta)
\label{FI}
\end{eqnarray}
where the one-particle spectrum close to the CP is again 
$\epsilon(p) = \sqrt{p^2+m^2}$ and the integral is computed like the 
bosonic one. The small-$x$ expansion yields 
\begin{eqnarray}
{C(x)\over {2\pi}} = -{\zeta(2)\over 4\pi^2} - 
{x^2\over 2}\left(\ln{x\over 2}+\gamma-{1\over 2}\right) +\nonumber\\
\sum_{l=2}^{\infty} 
\left(\!\!\begin{array}{c}1/2\\l\end{array}\!\!\right) 
x^{2l} \left(1-2^{2\,l-1}\right) \zeta(2l-1).
\label{CI}
\end{eqnarray}
Now the specific entropy is
\begin{equation}
{S\over L} = {\pi\over 6\,\beta} - \beta\,{m^2\over 4\pi} 
+ {\rm O}(m^4).
\label{SI}
\end{equation}

On the cylinder, the relative entropy $S_{\rm rel}(\l,\beta)$ 
includes a finite-size contribution from $C$, but in general differs from  
the $1d$ absolute entropy $S$. 
Let us see if there is a relation between them 
for solvable models. We calculate the relative entropy 
for $W=-\ln Z/(\beta\,L) = F/L$. For the Gaussian model, 
\begin{eqnarray}
S_{\rm rel}(r,\beta) = W(r,\beta) - W(0,\beta) - 
r{\partial W(r,\beta)\over \partial r} = S_{\rm rel}(r)\nonumber\\
\mbox{}+{1\over \beta^2}\left(C- C(0) - r{\partial C\over \partial r}\right) 
= \frac{r}{8\pi} - {S\over 2L\beta} + {\pi\over 6\,\beta^2},   
\label{StoSG}
\end{eqnarray}
For the Ising model the relative entropy is related instead to the energy
\begin{eqnarray}
S_{\rm rel}(m,\beta) = W(m,\beta) - W(0,\beta) - 
m{\partial W(m,\beta)\over \partial m} =\nonumber\\ S_{\rm rel}(m)
\mbox{}+{1\over \beta^2}\left(C- C(0) - m{\partial C\over \partial m}\right) 
=\nonumber\\  -\frac{m^2}{8\pi}\,(1+\ln {m^2\over \Lambda^2})
- ({E\over L}-e_0) + {\pi\over 12\,\beta^2}. 
\label{StoSI}
\end{eqnarray}
Thus only for the Gaussian model ${\cal C}$ and 
$\tilde\C$ are closely related. In any event, for both 
Gaussian and Ising models it is easy to derive series expansions of 
${\cal C}$ or $\tilde\C$. 

The components of the stress tensor can also be calculated exactly for free models. 
Defining $\Theta := T_a^a$ and $T := T_{11}-T_{22}-2\,i\,T_{12}$
we obtain \cite{Br}
\begin{eqnarray}
\langle \Theta(0) \rangle = 
\pm{m^2\over 2\pi}\left(K_0(0)+ 2\sum_{n=1}^\infty(\pm)^n K_0(n\,m\,\beta)\right),\\
\langle T(0) \rangle = 
\pm{m^2\over 2\pi}\left(K_2(0)+ 2\sum_{n=1}^\infty(\pm)^n K_2(n\,m\,\beta)\right),
\label{T0}
\end{eqnarray}
with the same sign convention as before. 
The modified Bessel functions are divergent at zero, namely, $K_0(0)$ is logarithmic 
divergent and $K_2(0)$ is quadratically divergent. These are UV divergences, 
like the $\Lambda$ divergent terms of $W$.

Using the recursion relations satisfied by the Bessel functions we can 
write the free energy (\ref{FK}) or (\ref{FI}) as
\begin{eqnarray}
\beta\,{F\over L} &=& e_0\, \beta 
\mp{m^2\,\beta\over 2\pi} \sum_{n=1}^\infty(\pm)^n \left[K_2(n\,m\,\beta)-
K_0(n\,m\,\beta)\right]\nonumber\\ 
&=& -{\beta\over 2} \,\langle T(0)-\Theta(0) 
\rangle =\beta\,\langle T_{22}(0) \rangle,
\end{eqnarray}
showing its relation with the expectation values of the 
components of the stress tensor, which generalizes off the CP 
the standard relation \cite{BCN}. 
Notice that it implies a definite form for $e_0$, namely,
$e_0 = \pm{m^2\over 4\pi}\left(K_0(0) - K_2(0)\right)$,
to be compared with (\ref{Wexpa}).

Similarly, we can calculate
\begin{eqnarray}
{\partial W\over \partial r} = {\partial e_0\over\partial r} \pm 
{1\over 2\pi} \sum_{n=1}^\infty(\pm)^n K_0(n\,m\,\beta) =
{1\over 2\,r}\,\langle\Theta(0) \rangle, \\
{E\over L}  = e_0\pm{m^2\over 2\pi} 
\sum_{n=1}^\infty(\pm)^n \left[K_2(n\,m\,\beta)+ K_0(n\,m\,\beta)\right] 
\nonumber\\= {1\over 2}\,\langle T(0) +\Theta(0) \rangle =
\langle T_{11}(0) \rangle. 
\end{eqnarray}
The first equation is just a particular case of the expression of the 
derivative of $W$ with respect to $r$ as the expectation value of the
``crossover part" of the action \cite{I-OC}, since $\Theta$ is
proportional to it. Furthermore, 
\begin{equation}
{S(r,\beta)\over \beta\,L} =   \pm{r\over \pi} 
\sum_{n=1}^\infty(\pm)^n K_2(n\,m\,\beta) \nonumber\\ =
\langle T(0) \rangle.
\end{equation}
Notice that this off-critical equation relates $\langle T \rangle$ with
the entropy and therefore disagrees with the surmise in Ref.\ \cite{MM}, which
relates it with the free energy.

The relations between thermodynamic quantities and the expectation values of 
components of 
the stress tensor obtained above are not restricted to free-field models. 
To see this one must regard those expectation values 
as the response of $W$ to various geometrical transformations \cite{BCN}; 
namely, a dilation in the $x_2$ direction, $\delta_LF = (F/L)\,\delta L$ is given 
by $T_{22}$, a dilation in the $x_1$ direction, $\delta_\beta(\beta F) = 
E\,\delta\beta$ is given by $T_{11}$, a dilation in both directions, equivalent 
to a change of $r$, is given by $\Theta$ and a dilation in one direction plus 
a contraction in the other ({\em shear transformation}) is given by $T$, 
corresponding to $S/(L \beta) = 
\langle T_{11}\rangle - \langle T_{22}\rangle =\langle  T\rangle$. However, 
closed expressions for these quantities can only be obtained for free models.
For interacting models one can compute the finite size
quantities above with conformal perturbation theory or the
thermodynamic Bethe ansatz \cite{I}.

In conclusion, the $2d$ relative entropy grows 
with the coupling, according to a general theorem \cite{I-OC}, 
which in $2d$ can be expressed in terms of 
the correlator of $\Theta$'s, similarly to Zamolodchikov's theorem. 
Nevertheless, the function ${\cal C}$ extracted from the relative entropy 
is not obviously connected with Zamolodchikov's $c$ function. 
On the other hand, the quantum $1d$ entropy 
satisfies a different monotonicity theorem 
with respect to $\beta$ or, say, the temperature, 
of pure thermodynamic nature, which leads to another 
monotonic function, $\tilde{\cal C}= (\beta/m)\,\langle  T\rangle$. 
The former function is well defined in the UV region whereas the latter 
is well defined in the IR region. 
As has been remarked before, a function monotonic with the RG 
is not unique \cite{CapFriLa}.
Unlike Zamolodchikov's $c$ function, the quantities ${\cal C}$ or 
$\tilde\C$ have a clear physical origin. 
Therefore, the entropic monotonicity theorems proposed here 
constitute an interesting alternative formulation.

Let's make two comments on possible generalizations. 
A different approach uses the Wilson RG \cite{WRG}. It is customary 
to try to prove the monotonicity theorem for the free energy. 
However, we believe that in that approach it is also 
some entropy the appropriate monotonic function, according to 
arguments presented previously \cite{I-OC}. 
Finally, the results in this paper can be generalized to higher dimensions 
($d=3$ or 4) and we expect them to contribute to the efforts to 
find a higher dimensional version of Zamolodchikov's $c$ theorem.

\vspace{3mm}
I acknowledge hospitality at the Nuclear Research Institute of Dubna (Russia), 
where this work was started, and partial support under Grant PB96-0887. 
I thank D. O'Connor for early conversations and M.A.R. Osorio for 
conversations on finite temperature QFT.

\end{multicols}
\end{document}